\documentclass[pre, twocolumn, amsmath, amssymb, showpacs,superscriptaddress
]{revtex4}
\usepackage{bm}
\usepackage[mathcal]{eucal}
\usepackage{nicefrac}
\usepackage{graphicx}
\usepackage{color}

\renewcommand{\vec}[1]{\bm{#1}}
\newcommand{\uvec}[1]{\hat{\vec{#1}}}

\newcommand{\avr}[1]{\left\langle#1\right\rangle}

\newcommand{\Lv}{\mathcal{L}}

\begin{document}

\title{Correlation of spin and velocity in granular gases}
\date{\today}

\author{W. T. Kranz}
\affiliation{Max Planck Institute for Dynamics and Self Organization, Bunsenstr. 10, 37073 G\"ottingen, Germany}
\author{N. V. Brilliantov}
\affiliation{Department of Mathematics, University of Leicester, University Road, Leicester LE1 7RH UK}
\author{T. P\"{o}schel}
\affiliation{Universit\"at Bayreuth, Physikalisches Institut, 95440 Bayreuth, Germany}
\author{A. Zippelius}
\affiliation{Max Planck Institute for Dynamics and Self Organization, Bunsenstr. 10, 37073 G\"ottingen, Germany}
\affiliation{Institute of Theoretical Physics, University of G\"ottingen, Friedrich-Hund-Platz 1, 37077 G\"ottingen}

\begin{abstract}
  In a granular gas of rough particles the spin of a grain is
  correlated with its linear velocity. We develop an analytical theory
  to account for these correlations and compare its predictions to
  numerical simulations, using Direct Simulation Monte Carlo as well
  as Molecular Dynamics. The system is shown to relax from an
  arbitrary initial state to a quasi-stationary state, which is
  characterized by time-independent, finite correlations of spin and
  linear velocity. The latter are analysed systematically for a wide
  range of system parameters, including the coefficients of tangential
  and normal restitution as well as the moment of inertia of the
  particles. For most parameter values the axis of rotation and the
  direction of linear momentum are perpendicular like in a sliced
  tennis ball, while parallel orientation, like in a rifled bullet,
  occurs only for a small range of parameters. The limit of smooth
  spheres is singular: any arbitrarily small roughness unavoidably
  causes significant translation-rotation correlations, whereas for
  perfectly smooth spheres the rotational degrees of freedom are
  completely decoupled from the dynamic evolution of the gas.
\end{abstract}


\maketitle

\section{Introduction}
\label{sec:introduction}
Materials which are composed of macroscopic objects, i.e. granular
media, attract increasing scientific interest due to their
importance in nature and technology, e.g.
\cite{Goldhirsch_AnnRev2003,LevyKalman2001}. The latter may be
exemplified by transport and storage of sand, cereals, granular
chemicals, etc. the former--by avalanches, land slides, dust devils,
etc. Spectacular celestial objects, like planetary rings or
interstellar dust clouds, can serve as another example of natural
granular systems \cite{GreenbergBrahic:1984}. The granular matter
exists there in a gaseous state and exhibits many properties of a
common molecular gas, e.g.
\cite{PoeschelLuding:2001,PoeschelBrilliantovNLP:2003,Goldhirsch_AnnRev2003,BrilliantovPoeschelOUP}.
The main (and very important) difference of a granular gas from a
molecular gas is the {\it dissipative} nature of particle
interactions, which are describe by macroscopic mechanics of solids
rather than by a microscopic interaction potential. The
consequences of the dissipative interactions are quite substantial: A
spatially homogeneous state is unstable
\cite{McNamaraYoung:1992,GoldhirschZanetti:1993,BritoErnst:1998},
velocities are not distributed according to a Maxwell-Boltzmann
distribution
\cite{GoldshteinShapiro1:1995,NoijeErnst:1998,EsipovPoeschel:1995,BreyCuberoRuizMontero:1999,
  DeltourBarrat:1997,HuthmannOrzaBrito:2000,BrilliantovPoeschel:1998VDF,GoldNoskBarLNP:2003,PoeschelBrilliantovFormella}
and the diffusion or self-diffusion is anomalous
\cite{BreyRuizMonteroGarciaRojo:1999,BreyRuizMonteroCuberoGarcia:2000,BrilliantovPoeschel:1998d,
  SantosDufty_PRL2001,GarzoMontanero_PRE2004}. These properties of a
granular gas have been observed for the case of smooth particles, when
grain collisions do not affect their rotational motion. This is,
certainly, an oversimplified model, since real grains have a rough
surface and exchange rotational and translational
energy in collisions.

Real granular particles experience {\em frictional forces} when
colliding. Hence, a more adequate model takes into account the
rotational motion of particles and the exchange of rotational and
translational energy in
collisions \cite{Goldhirsch_AnnRev2003,BrilliantovPoeschelOUP,HuthmannZippelius:1997,GoldshteinShapiro1:1995,
  ZippeliusLNP:2001,GoldNoskBarL_JPC2005,Elperin1997,JenkinsRichman1985,LunSavage1987,
  JaegerLiuNagelWitten1990,Luding1995,JenkinsLouge,BardenhagenBrackbillSulsky2000,
  CafLudHer_EPL2002,MitaraiHayakawaNakanishi2002}. Dissipative
frictional gases exhibit additional unusual features which are not present in molecular gases. For instance,
equipartition between rotational and translational motion does not hold \cite{HuthmannZippelius:1997} and the
hydrodynamic description requires an additional field and a dynamic equation to account for its evolution
\cite{GoldNoskBarL_PRL2005,GoldNoskBarL_JPC2005}. Moreover, the rotational and translational motion of
particles are correlated as mentioned in a very implicit way in Appendix E of \cite{GoldNoskBarL_JPC2005} and
worked out in \cite{BPKZ}. In the present study we analyze the latter effect in detail.

In Sec \ref{sec:model} we introduce a model of frictional particles and the  observables of interest.
Subsequently in Sec. \ref{sec:analytical-theory} an approximate analytical theory is developed and in Sec.
\ref{sec:simulations} we briefly explain the simulation techniques. The main results are presented in Sec.
\ref{sec:results}, where we compare predictions of the analytical theory with data from simulations. The
emphasis lies on the correlations in the quasi-steady state, but we also briefly discuss the relaxation to
the steady state. The technical details of the calculations are given in the Appendix.

\section{Model and Observables}
\label{sec:model}

We consider a granular gas consisting of $N$ inelastic hard spheres of
radius $a$, mass $m$, and moment of inertia $I = qma^2$. Here the
dimensionless variable q is determined by the mass distribution within
the disc. The state of
the system is fully described by the particles' positions $\{\vec
r_i\}$, velocities $\{\vec v_i\}$, and angular velocities
$\{\vec\omega_i\}$ for $i = 1, \ldots, N$.  The particles move freely
in between {\em instantaneous} collisions, whereupon their linear and
angular velocities change according to the collision rule: the
relative velocity at the point of contact of colliding particles
is
\begin{equation}
  \label{eq:1}
  \vec g \equiv \vec v_1 - \vec v_2
  + a\uvec n\times(\vec\omega_1 + \vec\omega_2)\,,
\end{equation}
with $\uvec n \equiv \uvec n_{12} \equiv \left(\vec r_1 - \vec r_2\right)/\left|\vec r_1 -
  \vec r_2\right|$. The post-collisional (primed) velocity is related
to the pre-collisional one by
\begin{equation}
  \label{eq:2}
  \begin{aligned}
    \vec g^{\prime}\cdot\uvec n &= -\varepsilon_n\,\vec g\cdot\uvec n\\
    \vec g^{\prime}\times\uvec n &= \varepsilon_t\,\vec g\times\uvec n\, .
  \end{aligned}
\end{equation}
The coefficient of normal restitution is denoted by $\varepsilon_n$ with
$0\leq\varepsilon_n\leq 1$. The value $\varepsilon_n=0$ implies no relative
motion in the normal direction after the collision, whereas for
$\varepsilon_n=1 $ no dissipation of the normal component of the
relative motion occurs.
The coefficient of tangential restitution has two elastic limits, namely
$\varepsilon_t=1$ corresponding to smooth spheres and $\varepsilon_t=-1 $ corresponding to perfectly rough
(reflecting) collisions without loss of energy for the tangential
motion. For all other values energy is lost in the tangential
component.
In general, both coefficients of restitution, $\varepsilon_n$ and
$\varepsilon_t$, depend on the impact
velocity \cite{BrilliantovSpahnHertzschPoeschel:1994,GGcool_98,GranRospap_99,BeckerSchwagerPoeschel:2008}.

Together with the conservation of linear and angular momentum the
collision rule, Eq. \eqref{eq:2}, determines the post-collisional
velocities in terms of the pre-collisional ones:
\begin{equation}
  \label{eq:3}
  \begin{aligned}
    \vec v^{\prime}_1 = \vec v_1 - \vec\delta,\quad
    \vec\omega^{\prime}_1 = \vec\omega_1 + \frac{1}{qa}(\uvec n\times\vec\delta)\\
    \vec v^{\prime}_2 = \vec v_2 + \vec\delta,\quad
    \vec\omega^{\prime}_2 = \vec\omega_2 + \frac{1}{qa}(\uvec n\times\vec\delta)\,,
  \end{aligned}
\end{equation}
where $m\vec\delta$ denotes the exchange of linear momentum with
\begin{equation}
\label{eq:def_delta}
\vec\delta \equiv \eta_t\vec g + (\eta_n - \eta_t)(\uvec n\cdot\vec g)\uvec n \, ,
\end{equation}
\begin{equation}
\label{eq:def_etaeta} \eta_n \equiv \frac{1 + \varepsilon_n}{2} \, , \qquad \qquad \eta_t \equiv
\frac{q}{2} \,\frac{1-\varepsilon_t}{1+q} \, .
\end{equation}

In the present study we address only non-driven systems. Moreover, we
focus on the homogeneous cooling state (HCS) of a gas, which is
characterized by two time-dependent granular temperatures, one for the
translational and one for the rotational motion,
\begin{equation}
  \label{eq:5}
  T = \frac{m}{3N}\sum_{i=1}^N\vec v_i^2 \quad\text{and}\quad
  R = \frac{I}{3N}\sum_{i=1}^N\vec\omega_i^2.
\end{equation}
One generally observes that after a transient period the system reaches a quasi-stationary state where
$r\equiv R(t)/T(t)=\text{const.}$, that is, both temperatures decay with the same rate. In general, $r\ne 1$
so that equipartition is violated. The value of $r$ depends on the collision parameters as well as on the
moment of inertia  \cite{HuthmannZippelius:1997,LudingHuthmannMcNamaraZippelius:1998}.

In this paper we focus on the correlation between the axis of
rotation of a granular particle and the direction of its linear
velocity, which may be quantified by the angle $\theta_i$ between
the linear and rotational velocity,
\begin{equation}
 \cos{\theta}_i =  \frac{\vec v_i\cdot\vec\omega_i}{|\vec v_i||\vec\omega_i|}.
\end{equation}
All information on the angle is contained in the distribution
\begin{equation}
\label{angular_dist} f(\cos{\theta}) =\frac{1}{N}\sum_{i=1}^N \delta(\cos{\theta}-\cos{\theta}_i),
\end{equation}
In a molecular gas all values of $\cos{\theta}$ occur with equal
probability due to equipartition. In contrast for a granular gas we know
that equipartition is violated and we expect to observe deviations
from the equi-distribution.

Because of symmetry, the average of $\cos{\theta_i}$ over all particles vanishes. Thus, a measure of
correlations is the second moment,
\begin{equation}
  \label{eq:4}
  \avr{\cos^2\theta} = \frac1N\sum_i
  \frac{(\vec v_i\cdot\vec\omega_i)^2}{\vec v_i^2\vec\omega_i^2}\,.
\end{equation}
If the angular and linear velocities are not correlated in their
direction, $\left < \cos^2 \theta \right>=1/3$. Hence, any deviation
of $\left < \cos^2 \theta \right>$ from $1/3$ indicates
correlations. Moreover, if $\left < \cos^2 \theta \right> < 1/3$ the
angular and linear velocities are preferably perpendicular, like in a
sliced tennis ball, while for $\left < \cos^2 \theta \right> > 1/3$
they are preferably aligned like in a rifled bullet.

\section{Analytical theory}
\label{sec:analytical-theory}

The evolution of any observable
\begin{equation}
F(t)=F\left(\left\{\vec r_i(t),\vec
    v_i(t),\vec\omega_i(t)\right\}\right)
\end{equation}
may be obtained by means of the pseudo-Liouville operator
$\Lv_+$ via
\begin{equation}
  \partial_t F(t)= i\Lv_+ F(t) \quad {\mbox {for}} \quad t>0.
\end{equation}
For hard spheres the pseudo-Liouville operator decomposes into two parts, $\Lv_+ = \Lv_0 + \Lv^{\prime}_+$, where
$\Lv_0 =\Lv_0^\text{tr}+ \Lv_0^\text{rot}$ describes the free
streaming of translational and rotational motion of particles. Here
$\Lv_0^\text{tr}= \sum_i\vec v_i\cdot \nabla_i$ and a similar expression
for $\Lv_0^\text{rot}$. The latter is not needed here,
because we never specify the orientation of our particles , which are
perfect spheres. The interaction part of the pseudo-Liouville operator reads,
$\Lv^{\prime}_+ = \sum_{i<j}\mathcal T_{ij}$, where the binary
collision operator $\mathcal T_{ij}$ reads
\cite{HuthmannZippelius:1997,NoijeErnst:1998}
\begin{equation}
  \label{eq:6}
  i\mathcal T_{ij} = -\uvec n_{ij}\cdot\vec v_{ij}
  \Theta\left(-\uvec n_{ij}\cdot\vec v_{ij}\right)\delta\left(r_{ij} - 2a\right)\left(\hat{b}_{ij} - 1\right) \, .
\end{equation}
The operator $\hat{b}_{ij}$ replaces unprimed by primed values
according to the collision rule, Eq.  \eqref{eq:3}. For example,
\begin{equation}
  \label{eq:b12_def}
  \hat{b}_{12} \vec{v}_1 = \vec{v}_1^{ \prime} \, , \qquad
  \hat{b}_{12} \vec{v}_2 = \vec{v}_2^{\prime} \, ,  \qquad
  \hat{b}_{12} \vec{v}_k = \vec{v}_k \, ,  \quad
  k \neq 1, \, 2
\end{equation}
with $\vec{v}_1^{ \prime}$ and  $\vec{v}_2^{\prime}$ given by Eq. \eqref{eq:3} and with similar relations for
the rotational velocities.

The ensemble average
of a dynamic variable is defined by
\begin{equation}
  \label{dyn_obs1} \avr F_t=\int d\Gamma \rho(0) F(t)=\int d\Gamma \rho(t) F(0)
\end{equation}
with $d\Gamma=\prod_{i}(d^3 r_i d^3 v_i d^3 \omega_i)$. Here
$F(t)=\exp{(-i\Lv_+t)}F(0)$ and
$\rho(t)=\exp{(-i\Lv_+^{\dagger}t)}\rho(0)$ denotes the $N$-particle
distribution, whose evolution is governed by the adjoint
$\Lv_+^{\dagger}$ of the evolution operator $\Lv_+$.
Differentiating Eq. \eqref{dyn_obs1} one obtains
\begin{eqnarray}
  \label{dyn_obs2} \frac{d}{dt}\avr F_t & = & \int d\Gamma \rho(0) \frac{d}{dt}F(t)
  =\int d\Gamma \rho(0) i \Lv_+ F(t)\nonumber\\
  & = &\int d\Gamma \rho(0) \exp{(i\Lv_+ t)}i \Lv_+ F(0)\nonumber\\
  & = & \int d\Gamma \rho(t)i \Lv_+ F(0)=\avr{i \Lv_+F}_t \, .
\end{eqnarray}

It is impossible to compute the time-dependent $N$-particle
distribution exactly, so that we have to resort to approximations. A
standard procedure in the analytical treatment of granular gases is to
assume homogeneity and molecular chaos, e.g.
\cite{BrilliantovPoeschelOUP} (see also \cite{MolChaos_03}). Under these assumptions the
$N$-particle velocity distribution function takes the form
\begin{equation}
  \label{fac_rho} \rho(t) =g_N( \vec{r}_1, \ldots, \vec{r}_N)\prod_i \rho_{1}(\vec v_i, \vec\omega_i,
  t) \, ,
\end{equation}
where the $N$-particle correlation function of a hard sphere system,
$g_N( \vec{r}_1, \ldots, \vec{r}_N)$, is not
affected by the particle roughness. For the HCS it may be approximated
by the corresponding function of an equilibrium hard-sphere system
(e.g.  \cite{BrilliantovPoeschelOUP}). For an isotropic system
$\rho_1(\vec v,\vec \omega)$ depends in general on $v=|\vec v|,
\omega=|\vec \omega|$ and the angle $\theta$ ($\cos \theta= \vec
v\cdot\vec \omega/\left(\left|\vec v\right| \, \left|\vec \omega\right|\right)$.
 Here we are
particularly interested in the dependence on $\cos\theta$ and expand
$\rho_1$ in Legendre polynomials $P_n(\cos\theta)$
\begin{equation}
  \label{eq:7}
  \begin{aligned}
    \rho_1(\vec v, \vec\omega, t) \propto
    &\exp\left(-\frac{m\vec v^2}{2T(t)}\right)
    \exp\left(-\frac{I\vec\omega^2}{2R(t)}\right)\\
    \times&
      \sum_{n=0}^\infty b_n(t)\vec v^n\vec\omega^nP_n(\cos\theta)\,,
      \end{aligned}
\end{equation}
where the $b_n(t)$ are time dependent expansion coefficients and the distribution function has to be
normalized according to $\int d\Gamma\rho_1 = 1$. We use a simple Gaussian even though the distributions are
non-Gaussian for strong dissipation and high densities. Deviations have been handled by an expansion in
Sonine polynomials \cite{goldhirschComputerAided}. Here we concentrate on the dependence on $\cos\theta$ and
leave a more general ansatz with both, angular correlations and non-Gaussian distributions, to future work.
To keep the calculations tractable, we limit the calculation to the lowest non-trivial order
\begin{equation}
  \label{eq:8}
  \begin{aligned}
    \rho_1(\vec v, \vec\omega, t) \propto
    &\exp\left(-\frac{m\vec v^2}{2T(t)}\right) \exp\left(-\frac{I\vec\omega^2}{2R(t)}\right)\\
    \times&\left[1 + b(t)\vec v^2\vec\omega^2P_2(\cos\theta)\right]\,,
  \end{aligned}
\end{equation}
where $b(t)\equiv b_2(t)$ and $P_2(\cos\theta) =
\nicefrac32(\cos^2\theta - \nicefrac13)$. The terms for odd $n$ vanish
by symmetry.

The lowest order coefficient $b(t)$ is simply related to the quantity
of interest $\avr{\cos^2\theta}_t$. Using $P_0(\cos\theta)=1$ and
expressing $\avr{\cos^2\theta}_t$ in terms of Legendre polynomials we
can write
\begin{equation}
  \label{eq:9}
  \begin{aligned}
    \avr{\cos^2\theta}_t = \frac13\int_v\int_\omega
    &[P_0(\cos\theta) + 2P_2(\cos\theta)]\\
    \times&[P_0(\cos\theta) + b(t)v^2\omega^2P_2(\cos\theta)] \,
  \end{aligned}
\end{equation}
where for brevity we introduce the shorthand notation
\begin{equation}
  \label{eq:shorthand}
\int_v  =\left(\frac{m}{2\pi T}\right)^{3/2} \int d^3 v \, \exp\left(-\frac{m v^2}{2 T}\right)
\end{equation}
and similarly for $\int_{\omega}.$ The angular integration in the Eq. \eqref{eq:9} may be performed using the
orthogonality relation for Legendre polynomials, yielding
\begin{equation}
  \label{eq:11}
  \avr{\cos^2\theta}_t
  = \frac13 + b(t)\frac{6T(t)R(t)}{5qm^2a^2} \, .
\end{equation}
Hence, the correlations of interest manifest themselves through the
coefficient $b(t)$--the larger the coefficient, the more pronounced
are deviations from the value $\avr{\cos^2\theta}
=1/3$ of the uncorrelated case.

To summarize our analytical approach so far: The time dependent
$N$-particle distribution has been parametrised by three
time-dependent functions $T(t)$, $R(t)$ and $b(t)$, which have to be
calculated self-consistently. This is achieved by applying the general
equation \eqref{dyn_obs2} for the evolution of an observable to
$T(t)$, $R(t)$ and $b(t)$ and using our ansatz for $\rho(t)$, see
Eqs. (\ref{fac_rho},\ref{eq:8}). Even with all these simplifying
assumptions, the analytical calculations are rather cumbersome and all
the details of the calculation have been relegated to the Appendix.

\begin{figure}[tb]
  \includegraphics{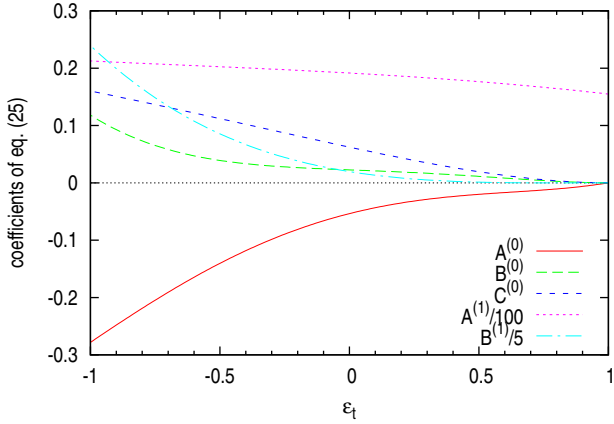}
  \caption{(Color online) The coefficients in Eq. \eqref{eq:A0_B1} as
    a function of $\varepsilon_t$ for $\varepsilon_n = 0.9$ and
    $q=2/5$. Note that $B^{(1)}$ is slightly negative for
    $\varepsilon_t\gtrsim 0.7$. Except for the coefficient $A^{(0)}$
    all coefficients vanish in the limit $\varepsilon_t\to 1$.}
  \label{fig:consts}
\end{figure}

The results are three first order differential equations for $T(t), R(t)$ and $b(t)$. These simplify, if we
measure times in units of the Enskog collision frequency $\omega_E=16(\pi T/m)^{1/2}na^2g_2(2a)$. In other
words we rescale time according to $d\tau = \omega_Edt$ and obtain:
\begin{equation}
  \label{eq:12}
  \begin{aligned}
    \frac{dT}{d\tau}
    = -AT(\tau) + B&\left[
      1 - \frac{b(\tau)}{2}\frac{T(\tau)R(\tau)}{qm^2a^2}
    \right]R(\tau)\\
    \frac{dR}{d\tau}
    = BT(\tau) - C&\left[
      1 - \frac{b(\tau)}{2}\frac{T(\tau)R(\tau)}{qm^2a^2}
    \right]R(\tau)
  \end{aligned}
\end{equation}
where
\begin{equation}
  \label{eq:13}
  \begin{aligned}
    A &\equiv \eta_n(1-\eta_n) + \eta_t(1 - \eta_t), \\
    B &\equiv \frac{\eta_t^2}{q} \, , \qquad
    C \equiv \frac{\eta_t}{q}\left(1 - \frac{\eta_t}{q}\right).
  \end{aligned}
\end{equation}
and
  \begin{multline}
    \label{eq:14}
      20\frac{db}{d\tau}
      = -b(\tau)\left[
        A^{(1)} + B^{(1)}\frac{R(\tau)}{T(\tau)}
        + \frac{40}{T(\tau)}\frac{dT}{d\tau}(\tau)\right.\\
\left.        + \frac{40}{R(\tau)}\frac{dR}{d\tau}(\tau)
      \right]\\
      - \frac{qm^2a^2}{T(\tau)R(\tau)}\left[
        A^{(0)} + B^{(0)}\frac{R(\tau)}{T(\tau)}
        + C^{(0)}\frac{T(\tau)}{R(\tau)}\right].
  \end{multline}
The constants are given by:
\begin{subequations}
    \label{eq:A0_B1}
    \begin{multline}
    \label{eq:15}
    A^{(0)} \equiv \frac{16}{3}\frac{\eta_t^3}{q}\left(\frac{2\eta_t}{q} - 1\right)
    - \frac23\frac{\eta_t^2}{q}\left(\frac{8\eta_t}{q} - 3\right)\\
    + \frac13\frac{\eta_t}{q}\left(\frac{\eta_t}{q} - 1\right)
    + \frac83\frac{\eta_t}{q}\left(\frac{\eta_t}{q} - 1\right)
    \eta_n(\eta_n - 1)
  \end{multline}
  \begin{align}
    \label{eq:16}
    B^{(0)} &\equiv \frac13\frac{\eta_t^2}{q}
    \left[\frac{16\eta_t}{q}\left(\frac{\eta_t}{q} - 1\right) + 5\right]\\
    C^{(0)} &\equiv \frac23\frac{\eta_t^2}{q}\left[
      8\eta_t(\eta_t - 1) +  4\eta_n(\eta_n - 1) + 3
    \right]
  \end{align}
  \begin{multline}
    \label{eq:17}
    A^{(1)} \equiv -\frac{8\eta_t^3}{q}\left(\frac{2\eta_t}{q} - 1\right)
    + \frac13\frac{\eta_t^2}{q}\left(\frac{24\eta_t}{q} - 37\right)\\
    - \frac56\frac{\eta_t}{q}\left(\frac{9\eta_t}{q} - 29\right)
    - \frac{4\eta_t\eta_n^2}{q}\left(\frac{\eta_t}{q} - 1\right)\\
    + \frac43\frac{\eta_t\eta_n}{q}\left(\frac{3\eta_t}{q} - 14\right)
    - 12\eta_t\eta_n\\
    + 22(\eta_t + \eta_n) - 6(\eta_t^2 + \eta_n^2)
  \end{multline}
  \begin{equation}
    \label{eq:18}
    B^{(1)} \equiv -\frac23\frac{\eta_t^2}{q}
    \left[\frac{8\eta_t}{q}\left(\frac{\eta_t}{q} - 1\right) + 1\right] \, .
  \end{equation}
\end{subequations}
Eqs. \eqref{eq:12} and \eqref{eq:14} constitute a set of
self-consistent equations for the observables $T(t), R(t)$, and
$b(t)$. Fig. \ref{fig:consts} illustrates the dependence of the above
coefficients on the coefficient of tangential restitution.

\section{Simulations}
\label{sec:simulations}

We performed both Direct Simulation Monte Carlo (DSMC) \cite{Bird:1994} and event-driven
Molecular Dynamics (MD) \cite{algo} calculations to check the
predictions of the analytical theory.
DSMC determines the stationary distribution of the scaled velocities
by numerically solving the kinetic Boltzmann equation which is based on the
assumption of molecular chaos. Consequently, for its application it is
assumed that the gas is uniform, thus, spatial correlations of the
particles are neglected. If this precondition is given, DSMC yields
very precise statistical results because of the large number of
particles which can be simulated (here we use $N=2\times 10^7$
particles\footnote{To be precise, although the mathematical operations
in DSMC looks like a particle simulation, the particles in the
simulation do not correspond to real particles. They are better
considered as quanta of probability \cite{BreyCubero:2000}.}).

Molecular Dynamics calculates the trajectories of the particles using
the collision rule, Eq. \eqref{eq:2}, therefore, MD allows to trace
the evolution of the correlation. On the other hand, MD is restricted
to much smaller systems as compared to DSMC. Although MD is free from
the mentioned assumptions, DSMC is significantly more efficient for a
homogeneous granular gas. Moreover, in the limit of low density both
methods provide, in principle, identical results for the stationary
state \cite{algo}. In practice, we use MD for $N=8000$ particles to
study the transient process of the system's relaxation to its
steady-state and up to $N=10^5$ for steady state correlations. The
volume fraction is $\frac{N}{V}\frac{4\pi a^3}{3}=0.0146$ or even
smaller, such that the gas is always in the HCS.

\section{Results}
\label{sec:results}

Starting from a random distribution of velocities and angular
velocities with mean $\left<\vec{v}\right>=\left<\omega\right>=0$,
after some transient period the system relaxes to a steady state where
the correlation of the spin and the translational velocity as well as
the ratio of translational and rotational temperatures adopt
stationary values.  We quantify these correlations by means of the
second moment $\left<\cos^2\theta\right>$, see Eq. \eqref{eq:4} and
analyze this quantity as a function of three parameters,
$\varepsilon_n$, $\varepsilon_t$, and $q$ in Sec. \ref{sec:steady}. The
relaxation to the steady state is discussed in Sec. \ref{sec:relax}
and in Sec. \ref{sec:beyond-second-moment} we consider correlations
beyond the second moment and investigate the distribution of
$\cos\theta$.

\subsection{Steady-state  correlations}
\label{sec:steady}

To study the
steady-state properties it is convenient to introduce an auxiliary
variable
\begin{equation}
  \label{eq:89}
  x(\tau) \equiv b(\tau) \frac{ T(\tau) R(\tau)}{q m^2 a^2}
  = \frac56 \left( \left<\cos^2\theta\right>_{\tau} - \frac13 \right) \, .
\end{equation}
Using $x(\tau)$ and $r(\tau)=R(\tau)/T(\tau)$ we recast the set of
three equations (\ref{eq:12},\ref{eq:14}) for $b$, $R$
and $T$ into a set of two equations for $x$ and $r$.  The result
reads
\begin{widetext}
  \begin{eqnarray}
  \label{eq:88}
    \frac{dr}{d\tau} &=& B - C\left[1 - \frac{x(\tau)}{2} \right]r(\tau) + Ar(\tau)
     - B\left[1 - \frac{x(\tau)}{2} \right]r^2(\tau) \\
    \label{eq:92}
    20\frac{dx}{d\tau} & = & -x(\tau)\left\{
      A^{(1)} + B^{(1)}r(\tau)
      - 20A - 20C + 20B[r(\tau) + r^{-1}(\tau)] + 20x(\tau)[C - B \, r(\tau)]/2
    \right\}  \\
    &-& A^{(0)} - B^{(0)}r(\tau) - C^{(0)}r^{-1}(\tau) \, . \nonumber
\end{eqnarray}
\end{widetext}

Setting the left hand side of Eqs. \eqref{eq:88} and \eqref{eq:92} to
zero one arrives at a set of coupled nonlinear equations for the stationary
values $r_{\infty} \equiv r(\tau\to\infty)$ and $x_{\infty} \equiv
x(\tau\to\infty)$. Instead of solving these equations directly, we
resort to an iteration scheme: At the outset we calculate a first
approximation of the temperature ratio $r^{(0)}_{\infty}$ neglecting
correlations, that is, for $x = 0$. Hence we assume that for moderate
inelasticity and roughness the temperature ratio is not noticeably
affected by the rotational-translational coupling. The result
reads
\begin{equation}
  \label{eq:93}
  r^{(0)}_{\infty} = \frac{A - C}{2B} + \sqrt{1 + \frac{(A - C)^2}{4B^2}} \, .
\end{equation}
Using this value for the stationary temperature ratio we then proceed
to calculate an approximate value of $x_{\infty}$
\begin{equation}
  \label{eq:94}
  x^{(0)}_{\infty} =
  - \frac{A^{(0)} + B^{(0)}r^{(0)}_{\infty} + C^{(0)}/r^{(0)}_{\infty}}
  {A^{(1)} + B^{(1)}r^{(0)}_{\infty} - 40C + 40B/r^{(0)}_{\infty}}
\end{equation}
where we use the fact, that
\begin{equation}
  \label{eq:95}
  B\left[r^{(0)}_{\infty} + 1/r^{(0)}_{\infty}\right]
  = A - C + 2B/r^{(0)}_{\infty}
\end{equation}
and neglect the terms quadratic in $x_{\infty}$ since they are
presumably small.  In principle, one could further iterate to get
better approximations, but we find that the results are reasonably
good already at this stage. For the more intuitive
variable, $\cos^2\theta$, Eq. \eqref{eq:94} implies
\begin{equation}
  \label{eq:96}
  \avr{\cos^2\theta}_{\infty} \approx \frac13
  - \frac65\frac{A^{(0)} + B^{(0)}r^{(0)}_{\infty} + C^{(0)}/r^{(0)}_{\infty}}
  {A^{(1)} + B^{(1)}r^{(0)}_{\infty} + 40B/r^{(0)}_{\infty} - 40C} \, .
\end{equation}

\begin{figure}[tb]
  \includegraphics[width=8cm]{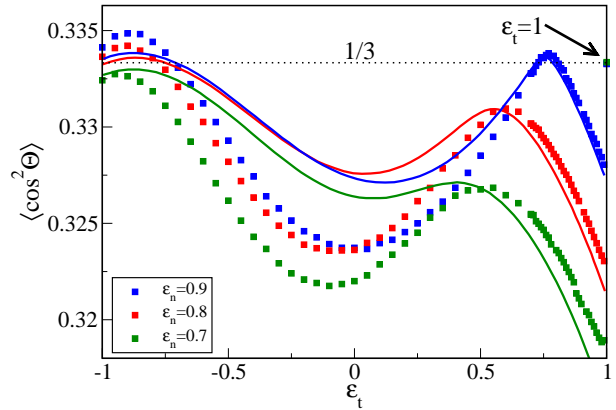}
  \caption{(Color online) Steady-state value of $\avr{\cos^2\theta}_{\infty}$
    as a function of the coefficient of tangential
    restitution, $\varepsilon_t$, for different $\varepsilon_n$.
    The predictions of the analytical theory, Eq. \eqref{eq:96},  are depicted by lines and
    points indicate the simulation data by DSMC. The line of vanishing
    correlations, $\avr{\cos^2\theta} =1/3 $ is shown, as well as the
    isolated point $\varepsilon_t=1$, which refers to the system of
    perfectly smooth hard spheres.  Note the existence of
    non-vanishing correlations even in the limit of smooth
    spheres, $\varepsilon_t \to 1$ (see Eq. \ref{eq:20}).}
  \label{fig:cos2beta}
\end{figure}

Fig. \ref{fig:cos2beta} shows the steady-state value of the correlation
factor $\avr{\cos^2\theta}_{\infty}$ as a function of
$\varepsilon_t$ for different values of $\varepsilon_n$ in comparison
with DSMC results.
Obviously, theory as well as simulations show that both types of
correlations may occur, $\avr{\cos^2\theta}<1/3$, as for a sliced
tennis ball or $\avr{\cos^2\theta}>1/3$ as for a rifled bullet
The dependence of the correlations on $\varepsilon_t$ is nonmonotonic with the strongest correlations for
$\varepsilon_t\sim 0$ and $\varepsilon_t\to 1$. Even though the dependence
on $\varepsilon_n$ is also not strictly monotonic, the dominant tendency
is an increase of correlations with decreasing
$\varepsilon_n$, i.e. increasing inelasticity.
The agreement between
theory and computer experiment is excellent for small inelasticity.
Moreover, even for significant dissipation the theory is able to
reproduce qualitatively the simulation results.

\begin{figure}[tb]
  \includegraphics{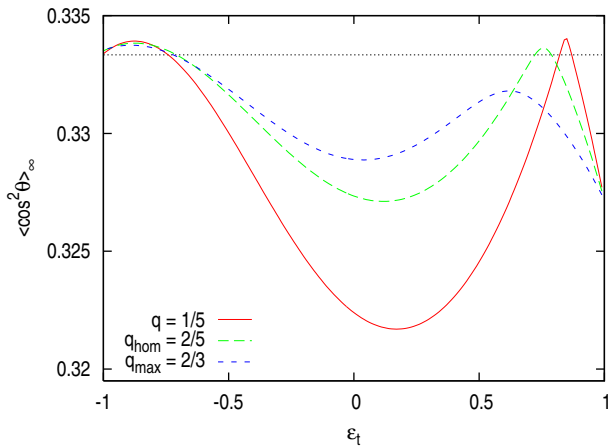}
  \caption{Steady-state value of
    $\avr{\cos^2\theta}_{\infty}$ for $\varepsilon_n = 0.9$ as a function
    of $\varepsilon_t$ and for different moments of inertia of a grain (see also
    Fig. \ref{fig:cos2qt}). With the decreasing moment of inertia the
    correlations become more sensitive to variations of the
    coefficient of tangential restitution}
  \label{fig:cos2q}
\end{figure}

Decreasing the moment of inertia, $q$, turns the magnitude of the correlations more sensitive
to changes in the coefficients of tangential restitution, as one can see from
Fig. \ref{fig:cos2q}. Interestingly, varying the moment of
inertia can even alter the type of the correlations: For instance, for
$q=1/5$ there exists a region for $\varepsilon_t >0$, where the rotation
axis is preferably directed along the linear velocity,
$\avr{\cos^2\theta}_{\infty}>1/3$, while for $q=2/3$ there is no such region.

\begin{figure}[tb]
  \includegraphics{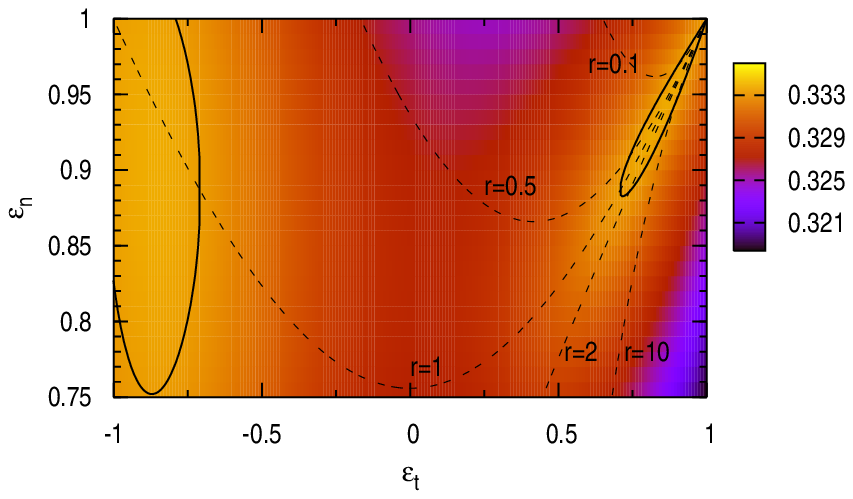}
  \includegraphics{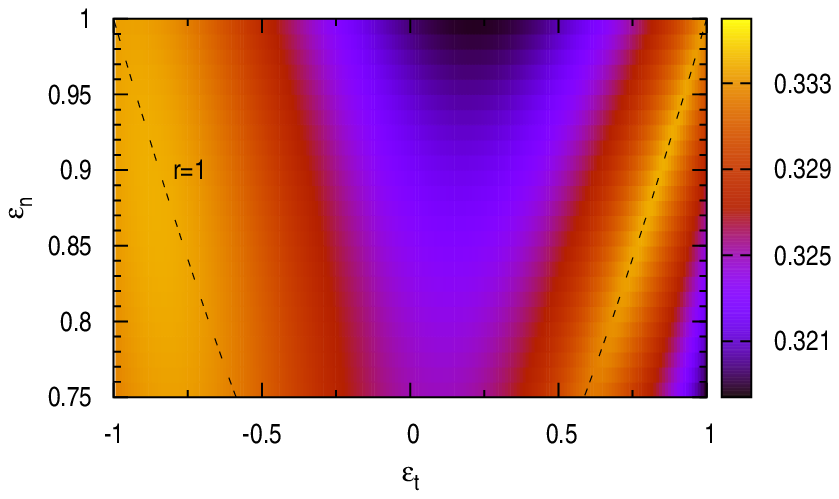}
  \includegraphics{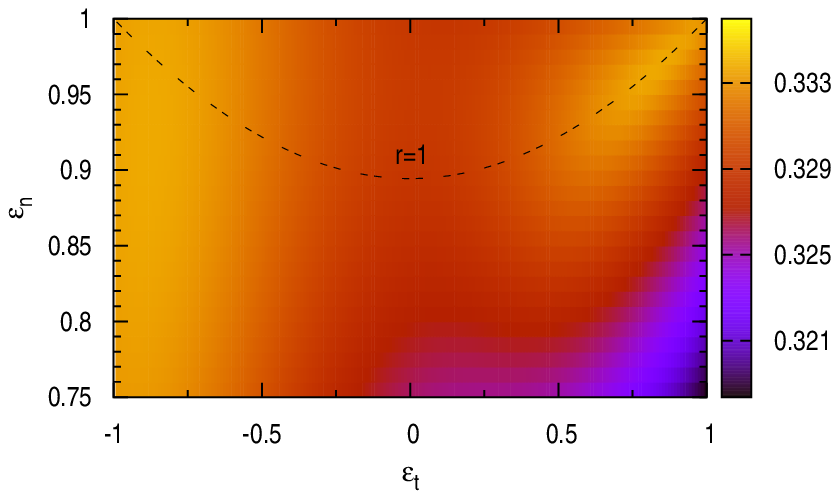}
  \caption{Stationary value $\avr{\cos^2\theta}_{\infty}$ (color
    coded) as a function of normal ($\varepsilon_n$) and tangential
    ($\varepsilon_t$) coefficients of restitution. The stationary
    value of the temperature ratio $r$ is superimposed through the
    dashed contour lines. The solid lines indicate vanishing
    correlations ($\avr{\cos\theta}_{\infty} = 1/3$). The moment of
    inertia $q=2/5$ (upper panel) corresponds to homogeneous spheres.
    The middle and bottom panel show the same data for $q=1/5$ and
    $q=2/3$, respectively}
  \label{fig:eb.25}
\end{figure}

Fig. \ref{fig:eb.25} (upper panel) illustrates the analytical result, Eq. \eqref{eq:96}
for the whole range of parameters $\varepsilon_t$ and
$\varepsilon_n$.
Note that for the majority of values of the
coefficients, $\avr{\cos^2\theta}_{\infty} < \nicefrac13$, that is, in
most cases the axes tend to be perpendicular to each other. Only in
two small regions of the parameter space the axes are preferably
parallel. The correlations vanish only for combinations of
$\varepsilon_n$ and $\varepsilon_t$ indicated by full lines. Dashed
lines show curves of constant $r$. Strong correlations appear for
large deviations from equipartition. This is shown more clearly in the
middle and bottom panels of Fig. \ref{fig:eb.25} which demonstrate the
rather strong influence
of the moment of inertia $I$ on the correlation factor
$\avr{\cos^2\theta}_{\infty}$.

\begin{figure}[tb]
  \includegraphics{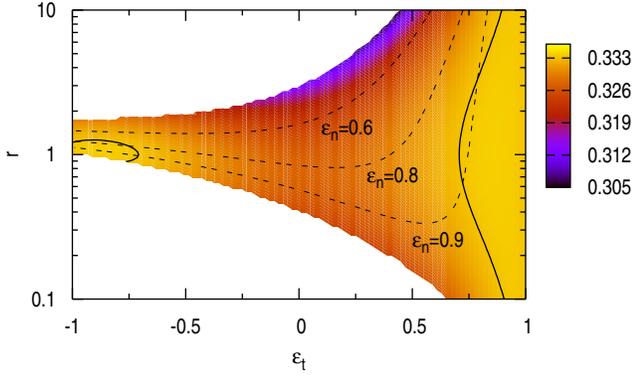}
  \caption{Stationary value of $\avr{\cos^2\theta}_{\infty}$ (color
    coded) as a function of the temperature ratio $r_{\infty}$ and the
    coefficient of tangential restitution $\varepsilon_t$. As previously, the solid lines
    indicate vanishing correlations and the dashed lines follow
    constant values of the coefficient of normal restitution
    $\varepsilon_n$. Note the logarithmic scale for the $r$-axis. The
    ragged border is an artifact of the limited numerical resolution}
  \label{fig:br.25}
\end{figure}

To check the assumption that strong correlations occur for strong
deviations from equipartition, we plot in Fig. \ref{fig:br.25} the
correlation factor $\avr{\cos^2\theta}_{\infty}$ as a function of
$r_{\infty}$ and $\varepsilon_t$. Technically this may be done, using
$\varepsilon_n = \varepsilon_n(r_{\infty})$--the inverse function of
$r_{\infty}=r_{\infty}(\varepsilon_n)$, given by Eq. \eqref{eq:93}, for
each fixed $\varepsilon_t$.  Note that
pronounced
correlations are present mainly for strong dissipation and large
temperature ratios. Also note the small range of admissible
temperature ratios for very rough spheres.

Analyzing Eq. \eqref{eq:96} in the limit of vanishing roughness,
\begin{equation}
  \label{eq:20}
  K^{(0)}\equiv\lim_{\varepsilon_t\to1}\avr{\cos^2\theta}_{\infty} - \frac13
  = -\frac{3}{8}\,\frac{1 - \varepsilon_n}{7 - \varepsilon_n}
\end{equation}
we see that even the smallest roughness induces finite correlations, for any given (fixed) value of the
coefficient of normal restitution, $\varepsilon_n\ne1$. For $\varepsilon_t = 1$, that is, for perfectly
smooth spheres, the initial rotational velocity of the particles is preserved. Therefore, the initial
rotational energy is preserved as well and $r$ does not reach a steady state. On the other hand
$\avr{\cos^2\theta}$ relaxes to the stationary value $\nicefrac13$ once the correlations in the initial
values of the translational velocities are lost due to collisions. Hence a straightforward expansion around
$\varepsilon_t = 1$ is problematic, or at least should be done with much care, as long as there is a finite
inelasticity $\varepsilon_n\ne1$. [See also the discussion of relaxation times in the following paragraph.]

\subsection{Relaxation to the steady-state}
\label{sec:relax}

So far we have discussed the quasi-stationary state, which is
characterized by constant $r$ and $\avr{\cos^2\theta}$. It is also
of interest to understand, how this stationary state is
reached---starting from arbitrary initial conditions.

Of particular interest is the limit of almost smooth spheres $\eta_t\propto \varepsilon_t-1\ll 1$ [see the
definition, Eq. \eqref{eq:def_etaeta}]. While the decay of the rotational temperature $R$ and the
translational temperature $T$ takes place extremely slowly, that is, with a rate $\sim \eta_t \ll 1$ [see Eq.
\eqref{eq:12} with $r=R/T \simeq A/B$ in this limit], the relaxation of the temperature ratio, $r=R/T$ as
well as of the correlation factor $x \sim (\avr{\cos^2\theta} -1/3)$ occurs on the collision time scale.
Indeed, in this limit one can write using Eqs. \eqref{eq:88}, \eqref{eq:92} and the definitions of the
coefficients (\ref{eq:13}) (\ref{eq:A0_B1}),
$$
d r /d \tau  \simeq -\eta_n(1-\eta_n ) \left( r/r_{\infty}^{(0)}\right) \left(r-r_{\infty}^{(0)} \right) \,,
$$
with $r_{\infty}^{(0)} \simeq A/B \sim 1/\eta_t^2 \gg 1$
from Eq. \eqref{eq:93}. This implies that $r$ relaxes to its stationary value exponentially fast with a rate
$\eta_n(1-\eta_n) = {\cal O}(1)$ (that is, on the collision time scale), while both temperatures $T$ and $R$
continue to decay with the same small rate.

\begin{figure}[tb]
  \includegraphics{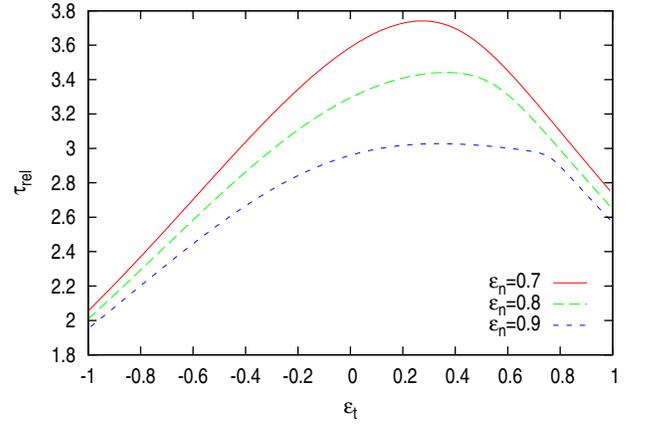}
  \caption{Relaxation time $\tau_\text{rel}$ (in the collision units)
    of $\avr{\cos^2\theta}_{\tau}$ when it approaches the steady-state
    value $\avr{\cos^2\theta}_{\infty}$. Note the narrow range of
    possible values for $\tau_{\rm rel}$}
      \label{fig:tau}
\end{figure}

To analyse the relaxation of $x(\tau)$ to its steady state value, we use
Eq. \eqref{eq:92} and approximate $r(\tau)$ by its steady state value
$r_{\infty}$:
  \begin{equation}
    \label{eq:97}
    \frac{dx}{d \tau} = -a_0 -a_1 x -a_2 x^2
  \end{equation}
where
\begin{eqnarray}
  \label{eq:97a}
  a_0 & =& \frac{1}{20} \left[  A^{(0)} + B^{(0)}r_{\infty} + C^{(0)}r^{-1}_{\infty}  \right]  \\
  a_1 & =& \left[ \frac{1}{20} A^{(1)} + \frac{1}{20} B^{(1)}r_{\infty} - A - C
    + B(r_{\infty} + r^{-1}_{\infty}) \right]  \nonumber\\
  a_2 & =&\frac12(C - Br_{\infty}) \nonumber \, .
\end{eqnarray}
The above equation with the initial condition $x(0)=0$ is solved by
\begin{equation}
  \label{eq:98}
  x(\tau) - x_{\infty} = - \frac{x_{\infty}}{1-\tanh \phi} \left[
    1-\tanh \left(\frac{\tau}{\tau_\text{rel}}+\phi\right)
  \right] \, ,
\end{equation}
with  the relaxation time
\begin{equation}
  \label{eq:rel_time}
  \tau_\text{rel} = \frac12 \sqrt{a_1^2 -4a_0 a_2}
\end{equation}
and $\tanh \phi =a_1/\sqrt{a_1^2 -4a_0 a_2}$. Evaluating the
coefficients for typical values of $\varepsilon_t$ and $\varepsilon_n$,
we find that the relaxation of the correlation factor
$\avr{\cos^2\theta}_t$ to its steady-state also occurs within a few
collisions per particle. This is illustrated in Fig. \ref{fig:tau},
where we plot the relaxation time $\tau_{\rm rel}$ given by Eq.
\eqref{eq:rel_time}.

We wish to stress here again, that the relaxation on the collisional
time scale to the steady state values applies only to the temperature
ratio and the mean square cosine of the angle between linear and
angular velocity. For nearly smooth particles, $\varepsilon_t \to 1$,
the relaxation of the rotational and translational temperatures is,
nevertheless, a very slow process, which proceeds with a small rate,
tending to zero as $\varepsilon_t \to 1$.

\begin{figure}[tb]
  \includegraphics{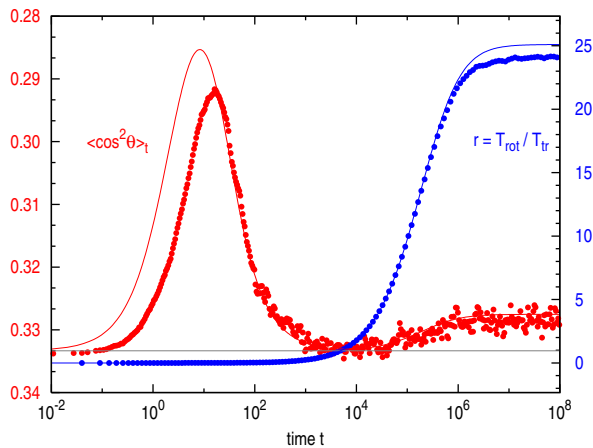}
  \caption{(Color online) Relaxation of $\avr{\cos^2\theta}_t$ and of the
    ratio of temperatures $r(t)=R(t)/T(t)$ to the steady state. Dots:
    molecular dynamics data for $8000$ particles, lines: analytical
    theory. To show that vanishing correlations $\avr{\cos^2\theta}_t$
    coincide with equipartition, we have chosen the vertical axes,
    such that the point $r=1$ on the right axis (blue) and the point
    $\avr{\cos^2\theta}_t=1/3$ on the left axis (red) have the same
    vertical height as indicated by a horizontal line.}
  \label{fig:dyn}
\end{figure}

To demonstrate the existence of several time regimes we discuss in the
following an instructive example. We initialize the particles with
$\vec{\omega}=0$ corresponding to $r=0$. The collision parameters are
$\varepsilon_n=\varepsilon_t=0.8$ so that the asymptotic value of the
ratio of temperatures is $r_\infty > 1$. We expect $r$ to
monotonically increase as a function of time---and this is indeed
observed as shown in Fig. \ref{fig:dyn}.
Now, we can check our hypothesis that correlations are small for values of $r$ close to equipartition. If the
hypothesis is correct, we should observe {\it non-monotonic} behavior of $\avr{\cos^2\theta}_t$. For short
times the correlations should be large and of tennis ball type, because grazing collisions are the most
effective for spinless particles to gain angular momentum. At intermediate times, when $r\sim 1$, the
correlations should be very small or vanishing. In the asymptotic state with $r_\infty > 1$, one should again
observe finite correlations.

These three time regimes are clearly born out in the time dependent
correlations, shown in Fig. \ref{fig:dyn}: (a) In the short time
regime ($0<t<10^3$) correlations are strong and $0<r<1$. (b) At
intermediate times ($10^3 <t < 10^5$) equipartition holds
approximately $r\approx 1$ and correlations are small or
vanishing. (c) The steady state ($t>10^6$) is characterized by $r \gg
1$ and finite $\avr{\cos^2\theta}_{\infty} <1/3$.
The agreement between analytical theory and molecular dynamics is good
also for the time-dependent quantities.
\begin{figure}[tb]
  \includegraphics{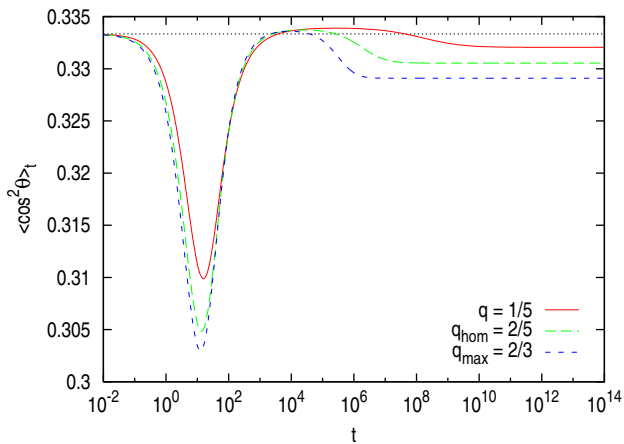}
  \caption{Impact of the grains' moments of inertia on the evolution
    and steady-state of $\avr{\cos^2\theta}_t$. The system parameters
    are $\varepsilon_n = 0.9$, $\varepsilon_t = 0.9$, $r(0) = 0.001$ and
    vanishing initial correlations. The values of $q$ ($I=qma^2$)
    represent spheres with the mass concentrated towards the center
    ($q = 1/5$), the homogeneous spheres ($q = 2/5$) and spheres with
    the mass concentrated mainly in the outer shell ($q=2/3$).}
  \label{fig:cos2qt}
\end{figure}

Fig. \ref{fig:cos2qt} demonstrates that the moment of inertia of the
particles does not change the evolution of
$\avr{\cos^2\theta}_t$ qualitatively. For the particular choice of the
coefficients of restitution the correlations are more pronounced for
larger $q=I/ma^2$ and fade with decreasing $q$. This however is not a
general rule; depending on the coefficients $\varepsilon_n$ and
$\varepsilon_t$, this tendency may reverse.

\begin{figure}[tb]
  \includegraphics{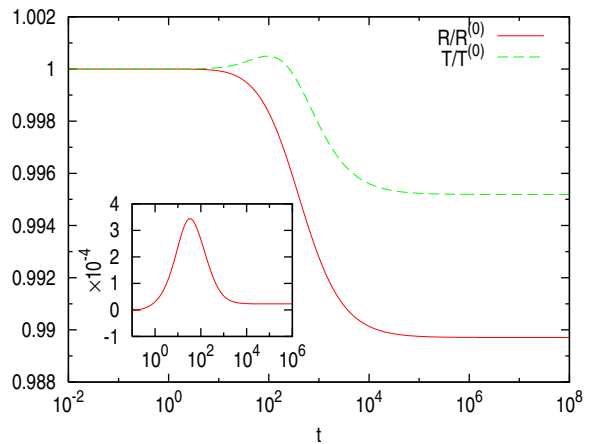}
  \caption{The ratio of the temperatures calculated with $b(t)$
    according to Eq. \eqref{eq:14} to those with $b(t)\equiv0$. The
    coefficients of restitution are $\varepsilon_n = \varepsilon_t = 0.8$,
    and the initial ratio of rotational to
    translational temperatures was set to the steady-state value $r(0)
    = r_{\infty}$. The inset shows $r(t)/r(0) - 1$ as a function of
    time. Note that the deviation of $r(t)$ from $r_{\infty}$ is
    always very small.}
  \label{fig:tno}
\end{figure}

The correlations between translational and rotational motion also have
a noticeable, albeit small impact on the basic characteristics of
granular gases---the translational and rotational temperatures. In
Fig.  \ref{fig:tno} we present the time dependence of
$R(t)/R^{(0)}(t)$---the ratio of the rotational temperature $R(t)$
with correlations to the corresponding value $R^{(0)}(t)$ without
correlations. The respective ratio $T(t)/T^{(0)}(t)$ for the
translational temperature is also plotted. Here we choose the case of
large $r \simeq 24$ ($\varepsilon_n = \varepsilon_t = 0.8$), which
correspond to $\avr{\cos^2\theta}_{\infty} < 1/3$, that is, for
preferably perpendicular rotational and translational velocity. Fig.
\ref{fig:tno} demonstrates that the effect of the correlations on the
granular temperatures $R(t)$ and $T(t)$ is indeed small. The
corresponding quantity $r(t)=R(t)/T(t)$ is also not sensitive to these
correlations. Moreover $r(t)$ does not deviate noticeably from its
steady-state value throughout the system's evolution, that is,
$|r(\tau)/r_{\infty}-1|\ll1$, as shown in the inset of
Fig.~\ref{fig:tno}.

\subsection{Beyond the second moment}
\label{sec:beyond-second-moment}

A complete one-particle picture includes the distribution
\begin{equation}
\label{full_dist}
 {\cal W}(\cos{\theta},v,\omega) =
  \frac{1}{N}\sum_{i=1}^N
\delta(\cos{\theta}-\cos{\theta}_i)\delta(v-v_i)\delta(\omega-\omega_i).
\end{equation}
Since correlations are developed in collisions, one intuitively expects that particles with larger
velocities, that  suffer stronger collisions, would show more pronounced orientational correlations; we
study these effects by binning the particles velocities.

\begin{figure}[tb]
  \includegraphics[width=0.9\columnwidth,clip]{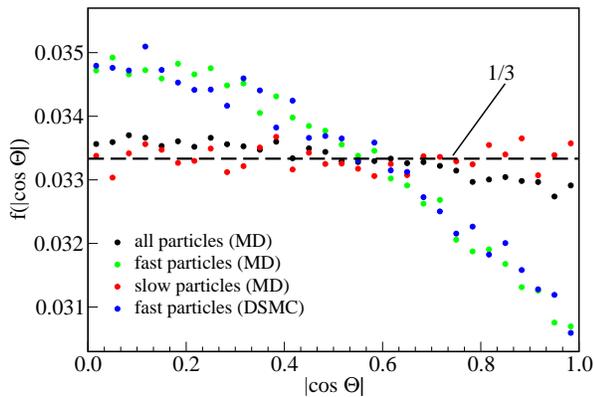}
   \caption{The angular distribution $f(|\cos \theta|)$ for the
    system of rough spheres with $\varepsilon_t=0.9$, $\varepsilon_n=0.9$
    and $q=2/5$ in the stationary state. Note that
    while there is no preferable angle between $\vec v$ and $\vec
    \omega$ for slow particles, correlations are clearly visible for
    fast particles favouring perpendicular linear and angular velocities.}
    \label{fig:cos}
\end{figure}

So far we discussed the correlation factor $\avr{\cos^2\theta}_{t}$, which is a second moment of the
distribution function ${\cal W}(\cos{\theta},v,\omega)$. Let us now analyze the distribution function itself.
Due to the limited statistics of our numerical data we discriminate only between two classes of particles:
the class of fast particles comprising those particles whose linear velocity belongs to the set of the 1/3
largest values {\em and } whose angular velocity belongs to the set of the 1/3 largest values. The class of
slow particles is defined correspondingly as the set of particles whose linear velocity belongs to the set of
the 1/3 smallest values {\em and} the angular velocity belongs to the set of the 1/3 smallest values.  In
Fig. \ref{fig:cos} we show the distributions $f (|\cos \theta|)$ for the two classes in comparison with the
distribution for all particles using both methods, MD and DSMC. In both cases we skipped the first 20
collisions per particle such that the ratio of temperatures, $r$, has reached its stationary value. For the
MD simulation we used a system of $N=10^5$ particles at low density (filling factor $<1\%$. Then we averaged
over 200 snapshots in distance of 1 collision per particle. In case of DSMC we used a system of $N=10^7$
particles and made the statistics based on a single snapshot. Both results agree very well.
The angular distribution is almost flat for slow particles and cannot
be distinguished from the distribution of all particles (within
statistical accuracy). On the other hand the fast particles exhibit a
nonuniform distribution with a maximum around $\cos \theta =0$.
Physically this means that the angle $\theta$ between $\vec v$ and
$\vec \omega$ for slow particles is uniformly distributed within the
interval $(0, \, \, \pi)$, while for fast particles it lies
preferentially around $\theta = \pi/2$. In other words, for the
particular choice of $\varepsilon_t=0.9$ and $\varepsilon_n=0.9$ the
fast particles tend to behave like sliced tennis balls, with $\vec
\omega$ perpendicular to $\vec v$.

\section{Conclusions and Outlook}
\label{sec:conclusion}
We have analysed in detail the correlations between rotational and
translational motion in a granular gas of frictional particles.  Under
the assumption of molecular chaos and homogeneity we have developed an
analytical theory which accounts for the correlations
$\avr{\cos^2\theta}_t$ in addition to the rotational $R(t)$ and
translational $T(t)$ temperature. We have also performed large
scale DSMC simulations as well as event
driven simulations to study the evolution
of a gas of rough spheres and in particular the above correlations.

We observe that the gas of {\it rough} particles always relaxes to a
steady-state with constant correlation $\avr{\cos^2\theta}_{\infty}$
and constant ratio $r_{\infty}=R(t)/T(t)$. While the relaxation of
$\avr{\cos^2\theta}$ and $r$ to their steady-state values happens on
the collisional time scale, the evolution of the rotational and
translational temperature in the near-smooth limit $\varepsilon_t \to
1$ is a slow process with a vanishingly small rate $\sim \eta_t \sim
(1- \varepsilon_t) \ll 1$. Physically, this may be explained as
follows. In the near-smooth limit the coupling of the rotational modes
to the translational ones becomes very weak. The energy of the
rotational motion of the particles is almost conserved in collisions
and the exchange of energy between the translational and rotational
degrees of freedom becomes very slow. Consequently the rotational
temperature as well as the translational temperature have a slowly
decaying component, governed by this weak exchange of
energies. However both temperatures decay with the {\it same} slow
timescale so that their ratio, $r$, is stationary - after it has
reached its steady state on the fast time scale of a few collisions.
Simultaneously, $\avr{\cos^2\theta }_t $ relaxes to its steady state
with a similar rate of the order of a few collisions.  We conclude
that the relaxation of the temperature ratio, $r$, and the angular
correlations is rapid, -- independent of the strength of the coupling
$(1- \varepsilon_t)$ as long as it is finite. Furthermore the
correlations persist up to a vanishingly small roughness and are
absent only for perfectly smooth particles,
$\varepsilon_t = 1$, which makes expansions around the smooth limit
questionable.

Our main results concern the correlation between the directions of
rotational and translational velocity in
the stationary state: The correlations depend sensitively on
the values of the coefficients of restitution and the moment of inertia;
for most of the system parameters
$\avr{\cos^2\theta} < 1/3$, implying that linear and angular
velocities are preferably orthogonal, like in a sliced tennis ball.
Only for a small part of the parameter space $\avr{\cos^2\theta} >
1/3$, which means that $\vec v$ and $\vec \omega$ are preferably
parallel like in a rifled bullet; the manifold of vanishing
correlations (in $\varepsilon_n, \varepsilon_t$ space) has seemingly zero
measure. The correlations are more pronounced for strong
deviations from equipartition.

Our approach can be extended in several directions. In the simulations
it is straightforward to use more advanced models for the coefficients
of restitution as functions of the impact velocity, e.g. \cite{BrilliantovSpahnHertzschPoeschel:1994,GGcool_98,GranRospap_99,BeckerSchwagerPoeschel:2008}.
It would also be of interest to study the full
one-particle distribution. Our results already indicate that more
energetic particles have stronger correlations, but a systematic study
has yet to be done. Furthermore, one expects to observe correlations
not only in very dilute gases, but also in rapidly moving denser
systems. Our approximate analytical theory is based on the assumption
of homogeneity and the density only enters into the Enskog collision
frequency, which sets the time scale. Hence our results for the
stationary state are independent of the density. This cannot hold true
in a rapidly moving dense system, yet we expect to observe
correlations as well. These could be analysed in a molecular dynamics
simulation either for a driven \cite{gayenOrientational} or undriven system.
 Finally, the observed
correlations may have important consequences for the stability theory
of dilute granular flows: they possibly alter the domain of stability
of granular system with respect to shear fluctuations---the main
instability of granular flows of smooth particles.

\paragraph*{Acknowledgement}

We thank Isaac Goldhirsch for interesting discussions; TK and AZ thank
Timo Aspelmeier for help with the MD simulations; TP acknowledges
support by a grant from G.I.F., the German-Israeli Foundation for
Scientific Research and Development.

\appendix*

\section{Analytical Calculations}
\label{sec:analyt-calc}

\subsection{Correlation factor}
\label{sec:calc-ddtavrd}
We present the details of the analytical
calculations, leading to the three self-consistent equations
\eqref{eq:12} and \eqref{eq:14} for
$T(t), R(t)$ and $b(t)$. First, we note that he computation of $b(t)$
or $\avr{\cos^2\theta}_t$ is severely hampered by the denominator in
Eq.~(\ref{eq:4}). Fortunately one can carry out the calculations with
the auxiliary observable
\begin{equation}
  \label{eq:30}
  \avr{\Delta}_t \equiv \frac{2}{3N}\sum_{i=1}^N
  \vec v_i^2\vec\omega_i^2P_2(\cos\theta_i) \, .
\end{equation}
Its relation to our set of observables can be established by
essentially the same steps as leading from Eq. \eqref{eq:9} to
Eq. \eqref{eq:11}:
\begin{equation}
  \label{eq:31}
  30b(t)\frac{T(t)R(t)}{qm^2a^2}
  = \avr{\Delta}_t\frac{qm^2a^2}{T(t)R(t)} \, .
\end{equation}
In the case of vanishing correlations we have $\avr{\Delta}_t = 0$.
Positive (negative) values correspond to a preference of a parallel
(perpendicular) orientation.

Owing to the assumptions of spatial homogeneity and molecular chaos it
suffices to consider the phase space of only a single pair of
particles (without loss of generality these shall be labeled $1$ and
$2$). Integrating out the spatial degrees of freedom and using the
definition of the pair correlation function
\begin{equation}
  \label{eq:def_g2}
  N(N-1) \int d\vec r_3 \ldots d \vec{r}_N g_N(\vec r_1, \ldots \vec r_N ) =n^2 g_2(r_{12}) \, ,
\end{equation}
with $n$ being the number density of the gas (e.g.
\cite{BrilliantovPoeschelOUP}) we obtain
\begin{equation}
  \label{eq:32}
  \begin{aligned}
    \avr{i\Lv_+ v\Delta}_t
    = &\nu \, N \, \int\limits_{v_1}\int\limits_{v_2}\int\limits_{\omega_1}\int\limits_{\omega_2}
    \left(\uvec n\cdot\vec v_{12}\right)\Theta\left(-\uvec n\cdot\vec v_{12}\right)\\[0.1cm]
    &\times\left[1 + b(t)\vec v_1^2\vec\omega_1^2P_2\left(\cos\theta_1\right)\right]\\[0.2cm]
    &\times\left[1 + b(t)\vec v_2^2\vec\omega_2^2P_2\left(\cos\theta_2\right)\right]\\[0.2cm]
    &\times\left(\hat b_{12} - 1\right)\Delta \,,
  \end{aligned}
\end{equation}
where $\uvec n$ is an arbitrary but fixed unit vector, $\nu = -8\pi
na^2 g_2(2a)$ and we used the shorthand notations
\begin{equation}
  \label{eq:33}
  \begin{split}
  \int_{v_i} &\equiv \left(\frac{m}{2\pi T}\right)^{3/2}\int d^3v_i
  \exp\left(-\frac{m\vec v_i^2}{2T}\right) \\
  \int_{\omega_i} &\equiv  \left(\frac{I}{2\pi R}\right)^{3/2}\int d^3\omega_i
  \exp\left(-\frac{I\vec \omega_i^2}{2R}\right) \, .
  \end{split}
\end{equation}
In the following we will drop the $b^2(t)$-term stemming from the
product of the two one particle distribution functions $\rho_1$ since
it was assumed to be small and we only want to go to first
order in $b(t)$.

The calculation of $\left(\hat b_{12} - 1\right)\Delta$ is obviously rather
involved and, thus, it needs to be broken up to stay tractable. It is
convenient to introduce relative integration variables
\begin{equation}
  \begin{split}
    \vec v \equiv \vec v_{12}/\sqrt2 ~~~~~& \vec V \equiv (\vec v_1 + \vec
    v_2)/\sqrt2\\
    \vec\omega \equiv \vec\omega_{12}/\sqrt2 ~~~~~& \vec\Omega \equiv (\vec\omega_1 +\vec\omega_2)/\sqrt2\,.
  \end{split}
\end{equation}
The term $\avr{i \Lv_+\Delta}_t$ can be broken
up along two different principles. First, one can make the dependence
on $b(t)$ explicit, that is,
\begin{multline}
  \avr{i\Lv_+\Delta}_t = \avr{\left(\hat  b_{12}-1\right)\Delta}^{(0)}\\ + b(t)\avr{\left(\hat b_{12}-1\right)\Delta}^{(1)} +\mathcal O\left(b^2\right)\,,
\end{multline}
where for any function $F$
\begin{equation}
  \label{eq:34}
  \avr{F}^{(0)}
  = \nu\int\limits_{v_1}\int\limits_{v_2}\int\limits_{\omega_1}\int\limits_{\omega_2}
  \left(\uvec n\cdot\vec v_{12}\right)\Theta\left(-\uvec n\cdot\vec v_{12}\right)\,F
\end{equation}
and
\begin{multline}
  \label{eq:35}
  \avr{F}^{(1)}
  = \nu\int\limits_{v_1}\int\limits_{v_2}\int\limits_{\omega_1}\int\limits_{\omega_2}
  \left(\uvec n\cdot\vec v_{12}\right)\Theta\left(-\uvec n\cdot\vec v_{12}\right)\\[0.1cm]
  \times\left[\vec v_1^2\vec\omega_1^2P_2\left(\cos\theta_1\right) +
  \vec v_2^2\vec\omega_2^2P_2\left(\cos\theta_2\right)\right]\,F \, .
\end{multline}
In order to be able to exploit some further symmetries it is advisable
to split up the last average again,
\begin{equation}
\avr{F}^{(1)} = \avr{F}^{\text{even}} + \avr{F}^{\text{odd}}
\end{equation}
where
\begin{multline}
  \label{eq:36}
    \avr{F}^{\text{even}}
    = \frac{3\sqrt2}{4}\nu\int\limits_v\int\limits_V\int\limits_\omega\int\limits_\Omega
    \left(\uvec n\cdot\vec v\right)\Theta\left(-\uvec n\cdot\vec v\right)\\
    \times\left[
    \left(\vec V\cdot\vec\Omega\right)^2 + \left(\vec
      V\cdot\vec\omega\right)^2 \right.\\
    + \left(\vec v\cdot\vec\Omega\right)^2 + \left(\vec v\cdot\vec\omega\right)^2\\
    -\left. \frac13\left(\vec V^2 + \vec v^2\right)\left(\vec\Omega^2 + \vec\omega^2\right)
    \right]\, F
\end{multline}
involves only even powers of $\vec V, \vec\omega, \vec\Omega$ and
\begin{multline}
  \label{eq:37}
    \avr{F}^{\text{odd}}
    = \frac{3\sqrt2}{2}\nu\int\limits_v\int\limits_V\int\limits_\omega\int\limits_\Omega
    \left(\uvec n\cdot\vec v\right)\Theta\left(-\uvec n\cdot\vec v\right)\\
    \times\left[
      \left(\vec V\cdot\vec\Omega\right)\left(\vec v\cdot\vec\omega\right)
      + \left(\vec V\cdot\vec\omega\right)\left(\vec
        v\cdot\vec\Omega\right) \right.\\
      - \left. \frac23\left(\vec V\cdot\vec v\right)\left(\vec\Omega\cdot\vec\omega\right)
    \right]F
\end{multline}
in contrast involves only the odd powers of these quantities.

Independently we can write
\begin{equation}
\Delta = \Delta_A - \Delta_B/3
\end{equation}
where
\begin{equation}
\Delta_A \equiv \sum_i\left(\vec
  v_i\cdot\vec\omega_i\right)^2~~~~\text{and}~~~~ \Delta_B \equiv \sum_i\vec v_i^2\vec\omega_i^2\,.
\end{equation}

First we address the $\Delta_A$-part. Applying the collision rule to
$\Delta_A$ yields
\begin{widetext}
  \begin{multline}
    \label{eq:38}
      \left(\hat b_{12} - 1\right)\Delta_A
      = \left(\vec\delta\cdot\vec\omega\right)^2 + \left(\vec\delta\cdot\vec\Omega\right)^2
      + \frac{1}{q^2a^2} \left[\left(\uvec n\times\vec\delta\right)\cdot\vec v\right]^2
      + \frac{1}{q^2a^2}\left[\left(\uvec n\times\vec\delta\right)\cdot\vec V\right]^2
      - \sqrt2\left(\vec\delta\cdot\vec\Omega\right)\left(\vec v\cdot\vec\Omega\right)\\
      - \sqrt2\left(\vec\delta\cdot\vec\Omega\right)\left(\vec V\cdot\vec\omega\right)
      - \sqrt2\left(\vec\delta\cdot\vec\omega\right)\left(\vec v\cdot\vec\omega\right)
      - \sqrt2\left(\vec\delta\cdot\vec\omega\right)\left(\vec V\cdot\vec\Omega\right)
      + \frac{\sqrt2}{qa}\left(\vec v\cdot\vec\Omega\right)
      \left(\uvec n\times\vec\delta\right)\cdot\vec v\\
      + \frac{\sqrt2}{qa}\left(\vec V\cdot\vec\omega\right)
      \left(\uvec n\times\vec\delta\right)\cdot\vec v
      + \frac{\sqrt2}{qa}\left(\vec v\cdot\vec\omega\right)
      \left(\uvec n\times\vec\delta\right)\cdot\vec V
      + \frac{\sqrt2}{qa}\left(\vec V\cdot\vec\Omega\right)
      \left(\uvec n\times\vec\delta\right)\cdot\vec V\\
      - \frac{2}{qa}\left(\vec\delta\cdot\vec\Omega\right)
      \left(\uvec n\times\vec\delta\right)\cdot\vec v
      - \frac{2}{qa}\left(\vec\delta\cdot\vec\omega\right)
      \left(\uvec n\times\vec\delta\right)\cdot\vec V
  \end{multline}
  and invoking the definition of $\vec\delta$, Eq.
  \eqref{eq:def_delta}, we obtain
  \begin{multline}
    \label{eq:39}
      \avr{\left(\hat b_{12}-1\right)\Delta_A}^{(0)}
      = 2\left(2\eta_t^2 - 2\eta_t + \frac{\eta_t^2}{q^2} - \frac{\eta_t}{q}
        + \frac{2\eta_t^2}{q}\right)
      \avr{\left(\vec v\cdot\vec\omega\right)^2}^{(0)}
      + 2\frac{\eta_t}{q}\left(\frac{\eta_t}{q} - 1\right)
      \avr{\left(\vec\omega\cdot\vec V\right)^2}^{(0)}\\
      + 2\left[2\left(\eta_n - \eta_t\right)^2 + \frac{\eta_t^2}{q^2}
        - \frac{2\eta_t}{q}\left(\eta_n - \eta_t\right)\right]
      \avr{\left(\uvec n\cdot\vec v\right)^2\left(\uvec n\cdot\vec\omega\right)^2}^{(0)}\\
      + 2\frac{\eta_t^2}{q^2}
      \avr{\left(\uvec n\cdot\vec V\right)^2\left(\uvec n\cdot\vec\omega\right)^2}^{(0)}
      + 2\eta_t^2a^2\avr{\left[\left(\uvec n\times\vec\Omega\right)\cdot\vec\omega\right]^2}^{(0)}
      + 2\frac{\eta_t^2}{q^2a^2}
      \avr{\left[\left(\uvec n\times\vec v\right)\cdot\vec V\right]^2}^{(0)}\\
      + 4\left[\left(2\eta_t - 1\right)\left(\eta_n - \eta_t\right)  - \frac{\eta_t^2}{q^2}
        + \frac12\frac{\eta_t}{q} - \frac{\eta_t^2}{q} +
        \frac{\eta_t}{q}\left(\eta_n -\eta_t\right)\right]
      \avr{\left(\uvec n\cdot\vec v\right)\left(\uvec n\cdot\vec\omega\right)
        \left(\vec v\cdot\vec\omega\right)}^{(0)}\\
      - \frac{2\eta_t}{q}\left(\frac{2\eta_t}{q} - 1\right)
      \avr{\left(\uvec n\cdot\vec V\right)\left(\uvec n\cdot\vec\omega\right)
        \left(\vec V\cdot\vec\omega\right)}^{(0)} \, .
  \end{multline}
  The terms that vanish by symmetry are already left out at this
  point.  The contributions to
  $\avr{\left(\hat b_{12}-1\right)\Delta_A}^{\text{even}}$ have exactly the same form.

  For $\avr{\left(\hat b_{12}-1\right)\Delta_A}^{\text{odd}}$ one finds the following
  contributions
  \begin{multline}
    \label{eq:40}
      \avr{\left(\hat b_{12}-1\right)\Delta_A}^{\text{odd}}
      = \left(\frac{4\eta_t^2}{q} - \frac{4\eta_t}{q} - 4\eta_t\right)
      \avr{\left(\vec v\cdot\vec\omega\right)\left(\vec V\cdot\vec\Omega\right)}^{\text{odd}}
      - \frac{4\eta_t}{q}\left(\eta_n -\eta_t\right)
      \avr{\left(\uvec n\cdot\vec v\right)\left(\uvec n\cdot\vec V\right)
        \left(\uvec n\cdot\vec\omega\right)\left(\uvec n\cdot\vec\Omega\right)}^{\text{odd}}\\
      + \frac{2\eta_t}{q}\left(1-2\eta_t\right)
      \avr{\left(\uvec n\cdot\vec V\right)\left(\uvec n\cdot\vec\Omega\right)
        \left(\vec v\cdot\vec\omega\right)}^{\text{odd}}
      - \frac{4\eta_t^2}{q}
      \avr{\left[\left(\uvec n\times\vec v\right)\cdot\vec V\right]
        \left[\left(\uvec n\times\vec\Omega\right)\cdot\vec\omega\right]}^{\text{odd}}\\
      + \left[4\left(\frac{\eta_t}{q} - 1\right)\left(\eta_n - \eta_t\right)
        + \frac{2\eta_t}{q}\right]
      \avr{\left(\uvec n\cdot\vec v\right)\left(\uvec n\cdot\vec\omega\right)
        \left(\vec V\cdot\vec\Omega\right)}^{\text{odd}} \, .
  \end{multline}
  Correspondingly, the $\Delta_B$-part may be written as
  \begin{multline}
    \label{eq:41}
      \avr{\left(\hat b_{12}-1\right)\Delta_B}^{(0)}
      = \frac{2\eta_t}{q}\left(\frac{\eta_t}{q} - 1\right)\left(2\eta_t - 1\right)^2
      \avr{\vec v^2\left(\vec n\times\vec\Omega\right)^2}^{(0)}
      + \frac{2\eta_t}{q}\left(\frac{\eta_t}{q} - 1\right)
      \avr{\vec V^2\left(\vec n\times\vec\Omega\right)^2}^{(0)}\\
      + \frac{2\eta_t^2}{q^2a^2}\left(2\eta_t - 1\right)^2
      \avr{\vec v^2\left(\uvec n\times\vec v\right)^2}^{(0)}
      + \frac{2\eta_t^2}{q^2a^2}\avr{\vec V^2\left(\uvec n\times\vec v\right)^2}^{(0)}
      + 4\eta_t\left(\eta_t - 1\right)\avr{\vec v^2\vec\omega^2}^{(0)}\\
      + 4\left(\eta_n^2 - \eta_n - \eta_t^2 + \eta_t\right)
      \avr{\left(\uvec n\cdot\vec v\right)^2\vec\omega^2}^{(0)}
      + \frac{8\eta_t}{q}\left(\frac{\eta_t}{q} - 1\right)
      \left(\eta_n^2 - \eta_n - \eta_t^2 + \eta_t\right)
      \avr{\left(\uvec n\cdot\vec v\right)^2\left(\uvec n\times\vec\Omega\right)^2}^{(0)}\\
      + \frac{8\eta_t^2}{q}\left(\frac{2\eta_t}{q} - 1\right)\left(2\eta_t - 1\right)
      \avr{\left[\left(\uvec n\times\vec v\right)\cdot\vec\Omega\right]^2}^{(0)}
      + \frac{8\eta_t^2}{q^2a^2}\left(\eta_n^2 - \eta_n - \eta_t^2 + \eta_t\right)
      \avr{\left(\uvec n\cdot\vec v\right)^2\left(\uvec n\times\vec v\right)^2}^{(0)}\\
      + 2\eta_t^2a^2\avr{\left(\uvec n\times\vec\Omega\right)^2\vec\omega^2}^{(0)}
      + 2\eta_t^2a^2\avr{\left(\uvec n\times\vec\Omega\right)^2\vec\Omega^2}^{(0)}
      + \frac{8\eta_t^3}{q}a^2\left(\frac{\eta_t}{q} - 1\right)
      \avr{\left(\uvec n\times\vec\Omega\right)^4}^{(0)}\\
      + \frac{8\eta_t^4}{q^2}
      \avr{\left(\uvec n\times\vec\Omega\right)^2\left(\uvec n\times\vec v\right)^2}^{(0)} \, .
  \end{multline}
  The contributions to $\avr{\left(\hat b_{12}-1\right)\Delta_B}^{\text{even}}$ again
  are formally equivalent to the above expression. This leaves us with
  \begin{multline}
    \label{eq:42}
      \avr{\left(\hat b_{12}-1\right)\Delta_B}^{\text{odd}} =
      -\frac{4\eta_t}{q}\left(2\eta_t - 1\right)
      \avr{\left(\uvec n\cdot\vec\omega\right)\left(\uvec n\cdot\vec\Omega\right)
        \left(\vec v\cdot\vec V\right)}^{\text{odd}}
      + \frac{8\eta_t^2}{q}
      \avr{\left[\left(\uvec n\times\vec v\right)\cdot\vec\omega\right]
        \left[\left(\uvec n\times\vec\Omega\right)\cdot\vec V\right]}^{\text{odd}}\\
      - \frac{8\eta_t}{q}\left(\eta_n - \eta_t\right)
      \avr{\left(\uvec n\cdot\vec v\right)\left(\uvec n\cdot\vec V\right)
        \left(\uvec n\cdot\vec\omega\right)\left(\uvec n\cdot\vec\Omega\right)}^{\text{odd}} \, .
  \end{multline}
\end{widetext}

We have now reduced the problem to the tedious but straightforward
calculation of a considerable number of averages. This task is best
suited for a computer algebra system and thus we only tabulate the
results. To simplify the notation we introduce the abbreviations $\tilde\nu \equiv \nu\sqrt{T/m\pi}$, $\tilde T \equiv T/m$, and
$\tilde R \equiv R/I$
\begin{subequations}
  \begin{equation}
    \label{eq:43}
    \avr{\left(\vec v\cdot\vec\omega\right)^2}^{\left(0\right)}
    = -4\tilde\nu\tilde T\tilde R
  \end{equation}
  \begin{equation}
    \label{eq:44}
    \avr{\left(\vec V\cdot\vec\omega\right)^2}^{\left(0\right)}
    = -3\tilde\nu\tilde T\tilde R
  \end{equation}
  \begin{equation}
    \label{eq:45}
    \avr{\left(\uvec n\cdot\vec v\right)^2\left(\uvec n\cdot\vec\omega\right)^2}^{\left(0\right)} =
    -2\tilde\nu\tilde T\tilde R
  \end{equation}
  \begin{equation}
    \label{eq:46}
    \avr{\left(\uvec n\cdot\vec V\right)^2\left(\uvec n\cdot\vec\omega\right)^2}^{\left(0\right)} =
    -\tilde\nu\tilde T\tilde R
  \end{equation}
  \begin{equation}
    \label{eq:47}
    \avr{[\left(\uvec n\times\vec\Omega\right)\cdot\vec\omega]^2}^{\left(0\right)} =
    -2\tilde\nu\tilde R^2
  \end{equation}
  \begin{equation}
    \label{eq:48}
    \avr{[\left(\uvec n\times\vec v\right)\cdot\vec V]^2}^{\left(0\right)} =
    -2\tilde\nu\tilde T^2
  \end{equation}
  \begin{equation}
    \label{eq:49}
    \avr{\left(\uvec n\cdot\vec v\right)\left(\uvec n\cdot\vec\omega\right)
      \left(\vec v\cdot\vec\omega\right)}^{\left(0\right)} =
    -2\tilde\nu\tilde T\tilde R
  \end{equation}
  \begin{equation}
    \label{eq:50}
    \avr{\left(\uvec n\cdot\vec V\right)\left(\uvec n\cdot\vec\omega\right)
      \left(\vec V\cdot\vec\omega\right)}^{\left(0\right)}
    = -\tilde\nu\tilde T\tilde R
  \end{equation}
  \begin{equation}
    \label{eq:51}
    \avr{\vec v^2\left(\uvec n\times\vec\Omega\right)^2}^{\left(0\right)} =
    -8\tilde\nu\tilde T\tilde R
  \end{equation}
  \begin{equation}
    \label{eq:52}
    \avr{\vec V^2\left(\uvec n\times\vec\Omega\right)^2}^{\left(0\right)}
    = -6\tilde\nu\tilde T\tilde R
  \end{equation}
  \begin{equation}
    \label{eq:53}
    \avr{\vec v^2\left(\uvec n\times\vec v\right)^2}^{\left(0\right)}
    = -12\tilde\nu\tilde T^2
  \end{equation}
  \begin{equation}
    \label{eq:54}
    \avr{\vec V^2\left(\uvec n\times\vec v\right)^2}^{\left(0\right)}
    = -6\tilde\nu\tilde T^2
  \end{equation}
  \begin{equation}
    \label{eq:55}
    \avr{\vec v^2\vec\omega^2}^{\left(0\right)}
    = -12\tilde\nu\tilde T\tilde R
  \end{equation}
  \begin{equation}
    \label{eq:56}
    \avr{\left(\uvec n\cdot\vec v\right)^2\vec\omega^2}^{\left(0\right)}
    = -6\tilde\nu\tilde T\tilde R
  \end{equation}
  \begin{equation}
    \label{eq:57}
    \avr{\left(\uvec n\cdot\vec v\right)^2\left(\uvec n\times\vec\Omega\right)^2}^{\left(0\right)}
    = -4\tilde\nu\tilde T\tilde R
  \end{equation}
  \begin{equation}
    \label{eq:58}
    \avr{[\left(\uvec n\times\vec v\right)\cdot\vec\Omega]^2}^{\left(0\right)}
    = -2\tilde\nu\tilde T\tilde R
  \end{equation}
  \begin{equation}
    \label{eq:59}
    \avr{\left(\uvec n\cdot\vec v\right)^2\left(\uvec n\times\vec v\right)^2}^{\left(0\right)}
    = -4\tilde\nu\tilde T^2
  \end{equation}
  \begin{equation}
    \label{eq:60}
    \avr{\left(\uvec n\times\vec\Omega\right)^2\vec\omega^2}^{\left(0\right)}
    = -6\tilde\nu\tilde R^2
  \end{equation}
  \begin{equation}
    \label{eq:61}
    \avr{\left(\uvec n\times\vec\Omega\right)^2\vec\Omega^2}^{\left(0\right)}
    = -10\tilde\nu\tilde R^2
  \end{equation}
  \begin{equation}
    \label{eq:62}
    \avr{\left(\uvec n\times\vec\Omega\right)^4}^{\left(0\right)}
    = -8\tilde\nu\tilde R^2
  \end{equation}
  \begin{equation}
    \label{eq:63}
    \avr{\left(\uvec n\times\vec\Omega\right)^2\left(\uvec n\times\vec v\right)^2}^{\left(0\right)}
    = -4\tilde\nu\tilde T\tilde R
  \end{equation}
\end{subequations}
\begin{subequations}
  \begin{equation}
    \label{eq:64}
    \avr{\left(\vec v\cdot\vec\omega\right)^2}^{\text{even}}
    = -24\tilde\nu\tilde T^2\tilde R^2
  \end{equation}
  \begin{equation}
    \label{eq:65}
    \avr{\left(\vec V\cdot\vec\omega\right)^2}^{\text{even}}
    = -15\tilde\nu\tilde T^2\tilde R^2
  \end{equation}
  \begin{equation}
    \label{eq:66}
    \avr{\left(\uvec n\cdot\vec v\right)^2\left(\uvec n\cdot\vec\omega\right)^2}^{\text{even}}
    = -6\tilde\nu\tilde T^2\tilde R^2
  \end{equation}
  \begin{equation}
    \label{eq:67}
    \avr{\left(\uvec n\cdot\vec V\right)^2\left(\uvec n\cdot\vec\omega\right)^2}^{\text{even}}
    = -3\tilde\nu\tilde T^2\tilde R^2
  \end{equation}
  \begin{equation}
    \label{eq:68}
    \avr{[\left(\uvec n\times\vec\Omega\right)\cdot\vec\omega]^2}^{\text{even}}
    = 2\tilde\nu\tilde T\tilde R^3
  \end{equation}
  \begin{equation}
    \label{eq:69}
    \avr{\left(\uvec n\cdot\vec v\right)\left(\uvec n\cdot\vec\omega\right)
      \left(\vec v\cdot\vec\omega\right)}^{\text{even}}
    = -12\tilde\nu\tilde T^2\tilde R^2
  \end{equation}
  \begin{equation}
    \label{eq:70}
    \avr{\left(\uvec n\cdot\vec V\right)\left(\uvec n\cdot\vec\omega\right)
      \left(\vec V\cdot\vec\omega\right)}^{\text{even}}
    = -6\tilde\nu\tilde T^2\tilde R^2
  \end{equation}
  \begin{equation}
    \label{eq:71}
    \avr{\vec v^2\left(\uvec n\times\vec\Omega\right)^2}^{\text{even}}
    = 6\tilde\nu\tilde T^2\tilde R^2
  \end{equation}
  \begin{equation}
    \label{eq:72}
    \avr{\vec V^2\left(\uvec n\times\vec\Omega\right)^2}^{\text{even}}
    = 3\tilde\nu\tilde T^2\tilde R^2
  \end{equation}
  \begin{equation}
    \label{eq:73}
    \avr{\left(\uvec n\cdot\vec v\right)^2\left(\uvec n\times\vec\Omega\right)^2}^{\text{even}}
    = 6\tilde\nu\tilde T^2\tilde R^2
  \end{equation}
  \begin{equation}
    \label{eq:74}
    \avr{[\left(\uvec n\times\vec v\right)\cdot\vec\Omega]^2}^{\text{even}}
    = 6\tilde\nu\tilde T^2\tilde R^2
  \end{equation}
  \begin{equation}
    \label{eq:75}
    \avr{\vec\omega^2\left(\uvec n\times\vec\Omega\right)^2}^{\text{even}}
    = 3\tilde\nu\tilde T\tilde R^3
  \end{equation}
  \begin{equation}
    \label{eq:76}
    \avr{\vec\Omega^2\left(\uvec n\times\vec\Omega\right)^2}^{\text{even}}
    = 7\tilde\nu\tilde T\tilde R^3
  \end{equation}
  \begin{equation}
    \label{eq:77}
    \avr{\left(\uvec n\times\vec\Omega\right)^4}^{\text{even}}
    = 8\tilde\nu\tilde T\tilde R^3
  \end{equation}
  \begin{equation}
    \label{eq:78}
    \avr{\left(\uvec n\times\vec v\right)^2\left(\uvec n\times\vec\Omega\right)^2}^{\text{even}} = 0
  \end{equation}
\end{subequations}
\begin{subequations}
  \begin{equation}
    \label{eq:79}
    \avr{\left(\vec v\cdot\vec\omega\right)\left(\vec V\cdot\vec\Omega\right)}^{\text{odd}}
    = -20\tilde\nu\tilde T^2\tilde R^2
  \end{equation}
  \begin{equation}
    \label{eq:80}
    \avr{\left(\uvec n\cdot\vec v\right)\left(\uvec n\cdot\vec\omega\right)
      \left(\vec V\cdot\vec\Omega\right)}^{\text{odd}}
    = -10\tilde\nu\tilde T^2\tilde R^2
  \end{equation}
  \begin{equation}
    \label{eq:81}
    \avr{\left(\uvec n\cdot\vec V\right)\left(\uvec n\cdot\vec\Omega\right)
      \left(\vec v\cdot\vec\omega\right)}^{\text{odd}}
    = -7\tilde\nu\tilde T^2\tilde R^2
  \end{equation}
  \begin{equation}
    \label{eq:82}
    \avr{\left(\uvec n\cdot\vec v\right)\left(\uvec n\cdot\vec V\right)
      \left(\uvec n\cdot\vec\omega\right)\left(\uvec n\cdot\vec\Omega\right)}^{\text{odd}}
    = -4\tilde\nu\tilde T^2\tilde R^2
  \end{equation}
  \begin{equation}
    \label{eq:83}
    \avr{[\left(\uvec n\times\vec v\right)\cdot\vec V]
      [\left(\uvec n\times\vec\Omega\right)\cdot\vec\omega]}^{\text{odd}} = 0
  \end{equation}
  \begin{equation}
    \label{eq:84}
    \avr{\left(\uvec n\cdot\vec\omega\right)\left(\uvec n\cdot\vec\Omega\right)
      \left(\vec v\cdot\vec V\right)}^{\text{odd}}
    = -2\tilde\nu\tilde T^2\tilde R^2
  \end{equation}
  \begin{equation}
    \label{eq:85}
    \avr{[\left(\uvec n\times\vec v\right)\cdot\vec\omega]
      [\left(\uvec n\times\vec\Omega\right)\cdot\vec V]}^{\text{odd}}
    = -5\tilde\nu\tilde T^2\tilde R^2 \, .
  \end{equation}
\end{subequations}
\vspace{0.2cm}

\subsection{The correction terms for the temperatures}
\label{sec:corr-terms-temp}

To calculate $dT/dt = \avr{i\Lv_+ T}_t$ one essentially proceeds along the same lines of reasoning
as detailed above. First of all, it is again advantageous to write the corrections to the
Gaussian distribution function explicitly, that is,
\begin{equation}
\avr{i\Lv_+T}_t = \avr{\left(\hat b_{12} -  1\right)T}^{(0)} + b(t)\avr{\left(\hat b_{12} - 1\right)T}^{(1)}
\end{equation}
where
\begin{multline}
  \label{eq:86}
    \frac{3}{4m}\left( \hat b_{12} - 1\right)T = \eta_t\left(\eta_t - 1\right)\left(\uvec n\times\vec v\right)^2\\
    + \eta_n\left(\eta_n - 1\right)\left(\uvec n\cdot\vec v\right)^2\\
    + \eta_t^2a^2\left(\uvec n\times\vec\Omega\right)^2\\
    + \eta_t\left(2\eta_t - 1\right)a\vec v\cdot\left(\uvec n\times\vec\Omega\right)
\end{multline}
and
\begin{equation}
  \label{eq:87}
  \begin{aligned}
    \frac{3}{4m}\left(\hat b_{12} - 1\right)R &= \frac{\eta_t^2}{q}\left(\uvec n\times\vec v\right)^2\\
    &+ \eta_t\left(\frac{\eta_t}{q} + 1\right)a^2\left(\uvec n\times\vec\Omega\right)^2\\
    &+ \eta_t\left(\frac{2\eta_t}{q} + 1\right)a
    \vec v\cdot\left(\uvec n\times\vec\Omega\right) \, .
  \end{aligned}
\end{equation}
The term $\avr{\left( \hat b_{12} - 1\right)T}^{(0)}$ is already known \cite{ZippeliusLNP:2001} and the
only other contribution is $\avr{\left(\uvec  n\times\vec\Omega\right)^2}^{\text{even}} = 2\tilde\nu\tilde T\tilde R^2$.

\bibliography{screws}

\end{document}